\documentclass[preprint,superscriptaddress,amsmath,amssymb,aps,pra]{revtex4-2}

\usepackage{graphicx}
\usepackage{dcolumn}
\usepackage{bm}
\usepackage{float}
\usepackage[colorlinks=true, allcolors=blue]{hyperref}

\begin{document}

\title{An ultra-stable three-dimensional photophoretic trap in air facilitated by a single multimode fiber}

\author{Souvik Sil}

\author{Anita Pahi}

\author{Aman Anil Punse}

\author{Ayan Banerjee}
\email{ayan@iiserkol.ac.in}
\affiliation{Department of Physical Sciences, IISER-Kolkata, Mohanpur 741246, India}

%\keywords{Optical Trapping, Photophoretic $\Delta T$ and $\Delta \alpha$ force, Multi-mode and Single-mode fiber, speckle pattern, instrumentation, methodology}

\date{\today}

% Abstract should be written in the present tense and impersonal style (i.e., avoid we), and be at most 200 words long
\begin{abstract}
Photophoretic forces - which are of thermal origin - have defined an alternative route of optical trapping of absorbing microparticles in air. Here, we show that a single multi-mode fiber facilitates significantly more robust optical traps compared to a pure Gaussian beam emanating from a single mode fiber for the trapping and manipulation of absorbing particles using photophoretic forces. We carefully study the dependency of trapping on speckle patterns generated from different modes from a multimode fiber, and experimentally observe that maximum trapping force can be obtained when the mean speckle size is comparable to the diameter of a trapped particle. We explain this observation by numerical simulations carried out to calculate the photophoretic force, and also determine stable trapping conditions from force balance equations. Interestingly, we also observe large oscillations of the trapped particle along the z-direction for multimode beams, which may be demonstrative of an effective restoring force for photophoretic trapping even in the axial direction. Our work may presage a new route for exciting applications on optical trapping and spectroscopy with photophoretic forces due to the inherent ease-of-use, portability, and flexibility of single muti-mode fiber based optical traps.
\end{abstract}
%\preprint{APS/123-QED}
\maketitle
\newpage
\section{Introduction}
Photophoretic forces - that arise due to the inhomogeneous heating of an absorbing particle in a gaseous medium by a intense light (such as a laser beam) - have introduced a new paradigm in optical trapping and manipulation of absorbing particles as demonstrated by a series of experiments\cite{shvedov_2009,desyatnikov2009,shvedov2010giant}. The photophoretic force\cite{horvath_2014} - which is also responsible for giant planet formation\cite{teiser2013aa} - can be categorized into two types of forces, $F_{\Delta T}$ and $F_{\Delta \alpha}$, where the former arises from an inhomogeneous temperature distribution across an absorbing particle's surface due to laser heating of that particle, and is directed from the hotter side to the colder side of the particle, while the later results from a difference in the accommodation coefficient of the particle, and measures the efficiency of heat exchange between the heated particle and ambient molecules of the medium\cite{horvath_2014}. Though it is clear that the magnitude of photophoretic forces depends on the intensity of light, it needs to be pondered whether the magnitude only depends on intensity or the gradient of intensity - especially if one uses structured light. In our earlier works\cite{bera_2016,sil_2017,sil2020study}, we showed that particles are confined in at least one dimension due to the photophoretic forces balancing the gravitational force, while in the other dimension, there exists a restoring force possibly generated due to the complex motion of the particle in the light field. Thus, due to the force balance, particles of a certain mass can be trapped with the appropriate intensity of the trapping light\cite{sil2022trapping}, with higher and lower intensities leading to particles escaping the trap\cite{sil_2017,sil2022trapping}. But in the presence of a structured intensity profile, i.e., adjacent patterns of light and dark regions - there could be opposite forces on a particle as it traverses the intensity profile. This would result in a complex restoring force on the particle even in the direction of gravity, and not merely a force balance - leading to a more efficient trap and a definite dependence of the trapping efficiency on the intensity profile of the trapping beam. The challenge, however, is to create such light beams having a complex intensity profile.

Several experimental configurations have been developed to achieve such intensity profiles. These include counter-propagating vortex beams\cite{desyatnikov2009}, doughnut-like vortex beam shapes created by two hollow beams overlapping in the trapping volume\cite{shvedov2010giant},optical lattices\cite{shvedov2012optical}, tapered rings\cite{liu_2014_photophoretic}, dark-hollow\cite{porfirev2015dark} and optical hollow-cone beams\cite{redding2015optical,lamhot2010self}, etc. Moreover, to increase  flexibility, holographic beam shaping has been utilized to form optical bottle beams, either by a Moire\' technique employing  spatial light modulators\cite{zhang2011trapping}, or by creating discrete trapping
sites\cite{alpmann2012holographic}, that lead to a vanishing intensity region surrounded by light in all three dimensions. Further, such structured light fields can be created using a speckle pattern\cite{goodman1976some}, that can be generated in various ways, such as the scattering of laser light from a rough surface or transverse mode-mixing in a multi-mode fiber. For example, Shvedov et al. demonstrated that a volume speckle field generated by a coherent laser beam and diffuser can be used to confine a massive numbers of carbon particles in air using photophoretic forces\cite{shvedov2010,shvedov2010laser}. However, the optical configurations to achieve such structured profiles are quite complex and have very small alignment tolerance. Recently, speckle optical tweezers (ST) have been developed where a speckle pattern, generated using a multi-mode fiber, has been used for performing collective optical manipulation of high-refractive-index particles\cite{volpe2014brownian}, and even controlled manipulation of high and low refractive index micro-particles and nano-particle loaded vesicles\cite{jamali2021speckle}. Besides, opto-thermoelectric speckle tweezers have been developed very recently where an optical speckle field was fed into a thermal speckle field through interaction with plasmonic substrates,  thus converting the high-intensity speckle grains into corresponding thermal speckle grains\cite{kotnala2020opto}.

The speckle optical tweezers described above have predominantly been employed in liquid  media and demonstrate trapping in two dimensions. In this paper, however, we report for the first time the use of a multi-mode fiber to create a photophoretic trap in air for trapping and manipulating mesoscopic absorbing particles in all three dimensions. It is important to note that speckle patterns generated from a multi-mode fiber have advantages of uniform speckle distribution, easy alignment, high optical transmission efficiency, and high flexibility. Besides, optical fiber traps provide substantial benefits over conventional microscope-based optical tweezers and are more advantageous than free-space photophoretic traps - the main advantages being large working distance and ease of alignment, which finally results in significantly increased convenience in trapping. In addition, we observe that a multi-mode beam profile exerts a radial trapping force that is about eight times stronger than that by a single-mode (Gaussian) beam profile. We also attain large manipulation velocities of around 5 $mm/s$, both axially and radially, for the trapped particles  using the multi-mode beam profile. Note that this velocity is presently limited by our experimental capabilities, and can possibly reach even higher values. We also determine the dependence of the trapping force on the nature of the speckle pattern, and show that the maximum trapping force is exerted when the average speckle size is similar to that of the particles. We  perform an analysis using the multiphysics tool COMSOL to explain our experimental observations.  Further, we also attempt to explain the origin of the large trapping force exerted by the speckle pattern compared to that by a Gaussian beam profile from a simple model based on a balance of all the forces that a trapped particle experiences. Our model predicts axial trajectories of the trapped particles with the different beam profiles which we compare with those measured in our experiments, and achieve reasonable agreement. 

\section{Materials and Methods}
Earlier, we demonstrated through a series of experiments that an absorbing particle could be trapped employing photophoretic forces, generated by a fundamental Gaussian beam in free space \cite{bera_2016,sil2020study}, or through a single-mode fiber\cite{sil2022trapping}. In these experiments, it was clear that in our experimental configuration, the particles are confined in the axial direction due to the photophoretic $\Delta T$ force, while in the radial direction, a restoring force appears to be generated by the helical motion of the particle caused by the transverse photophoretic body force ($F_{\Delta \alpha}$), which applies a torque on particles due to its interaction with gravity\cite{sil2020study,rohatschek_1995}. As a result of this, the particle trajectories are found to be radially shifted off-axis with respect to the trapping beam center\cite{bera_2016}, with the trap stiffness being linearly proportional to the laser power or intensity\cite{bera_2016,sil2020study}. Thus, it is apparent that any beam profile which has a large transverse extent resulting in a high off-axis intensity, would increase the trapping efficiency. This is indeed the case - as we showed experimentally - where the trapping efficiency due to a beam profile that was the superposition of a fundamental Gaussian and the first-excited state (Hermite-Gaussian mode) generated by a quasi-single mode optical fiber, was higher by around 80\% compared to just the fundamental Gaussian mode\cite{sil_2017}. Extrapolating from this observation, we considered coupling the trapping laser using a multi-mode fiber for our experiments. This was due to the fact that a multi-mode fiber has a higher mode volume than a single-mode one, giving even higher off-axis intensity compared to that we achieved in our previous experiments, and our intuition was that this would increase the trapping efficiency even further. Thus, in the first set of results we report, we quantify the trapping efficiency for input laser mode profiles generated by a multi-mode, and a single-mode fiber. We  determine the radial trapping force by the well-known viscous drag method\cite{bera_2016}, and also measure the threshold laser power for trapping.

In the experiments, we use a graded-index multi-mode optical fiber (Thorlabs GIF625) with core diameter 62 $\mu m$, around ten times higher mode volume compared to a single-mode fiber with a core diameter of 6 $\mu m$. The total number of guided modes (N) for a typical graded-index fiber can be defined as $N = \frac{q}{2(q+2)} V^2$\cite{ghatak1998introduction}, where q is the exponent of the power-law profile which has a value of 2 for typical graded-index multi-mode fibers, and V [$= K_0 a NA$] is the waveguide parameter - signifying, the number of linearly polarized  (LP) modes propagating through the fiber for a given wavelength, where $K_0$ ($= \frac{2\pi}{\lambda}$) is free space propagation constant, $a$ is the core radius, and $NA$ is the numerical aperture of the fiber\cite{ghatak1998introduction}. By putting the value of $K_0$ where $\lambda = 671 nm$,  $a (= 31.5 \mu m)$ and $NA (= 0.275)$, the value of V becomes 80 - signifying that the multi-mode fiber can support around 1200 LP modes. Thus, due to the superposition of all those LP modes, a completely random distribution of electric fields, which are typical termed as a speckle pattern - appears at the output of the fiber, which is shown in Fig.~\ref{Fig1}(a).

\begin{figure}[h]
	\centering
%	\captionsetup{justification=centerlast}
	\includegraphics[scale=0.52]{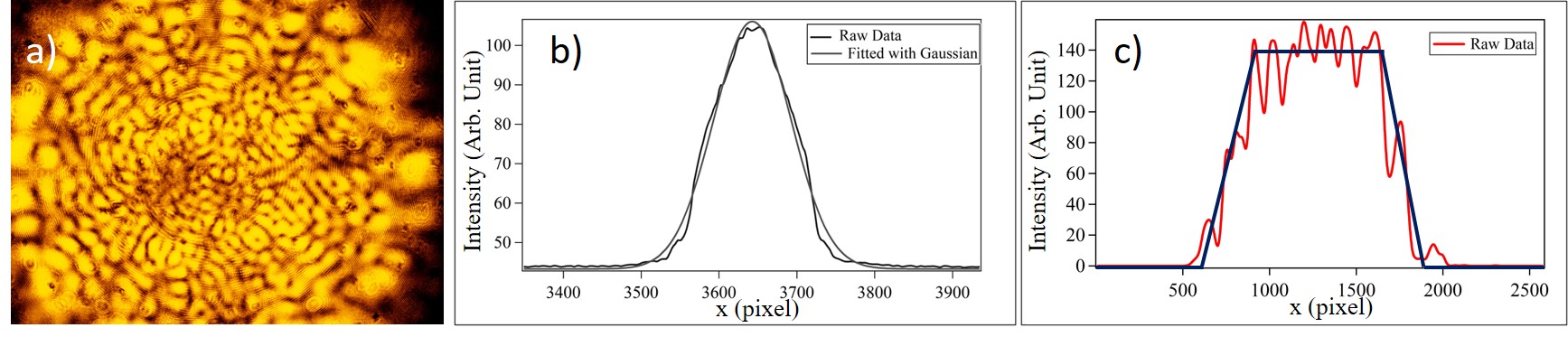}
	\caption{(a) Typical speckle pattern of Multimode fiber; (b) $\&$ (c)  1-D plot profile of Gaussian and Multimode beam profiles, respectively} 
	\label{Fig1}
\end{figure}    

The line plot of Gaussian and multi-mode beam profile for the same beam size are shown in Fig.~\ref{Fig1} (b) and (c) respectively, in which the raw data (black line) of Fig.~\ref{Fig1} b) is fitted with a standard Gaussian function (gray line). But for the multi-mode case, we can approximate the 1-D profile with a top hat function as shown in the blue line of Fig.~\ref{Fig1} c) - demonstrating the increased transverse extent, as well as higher off-axis intensity compared to the Gaussian beam profile. Our expectation was that this would increase the trapping efficiency for the multi-mode beam profile. \\

\begin{figure}[h]
	\centering
	%\captionsetup{justification=centerlast}
	\includegraphics[scale=0.47]{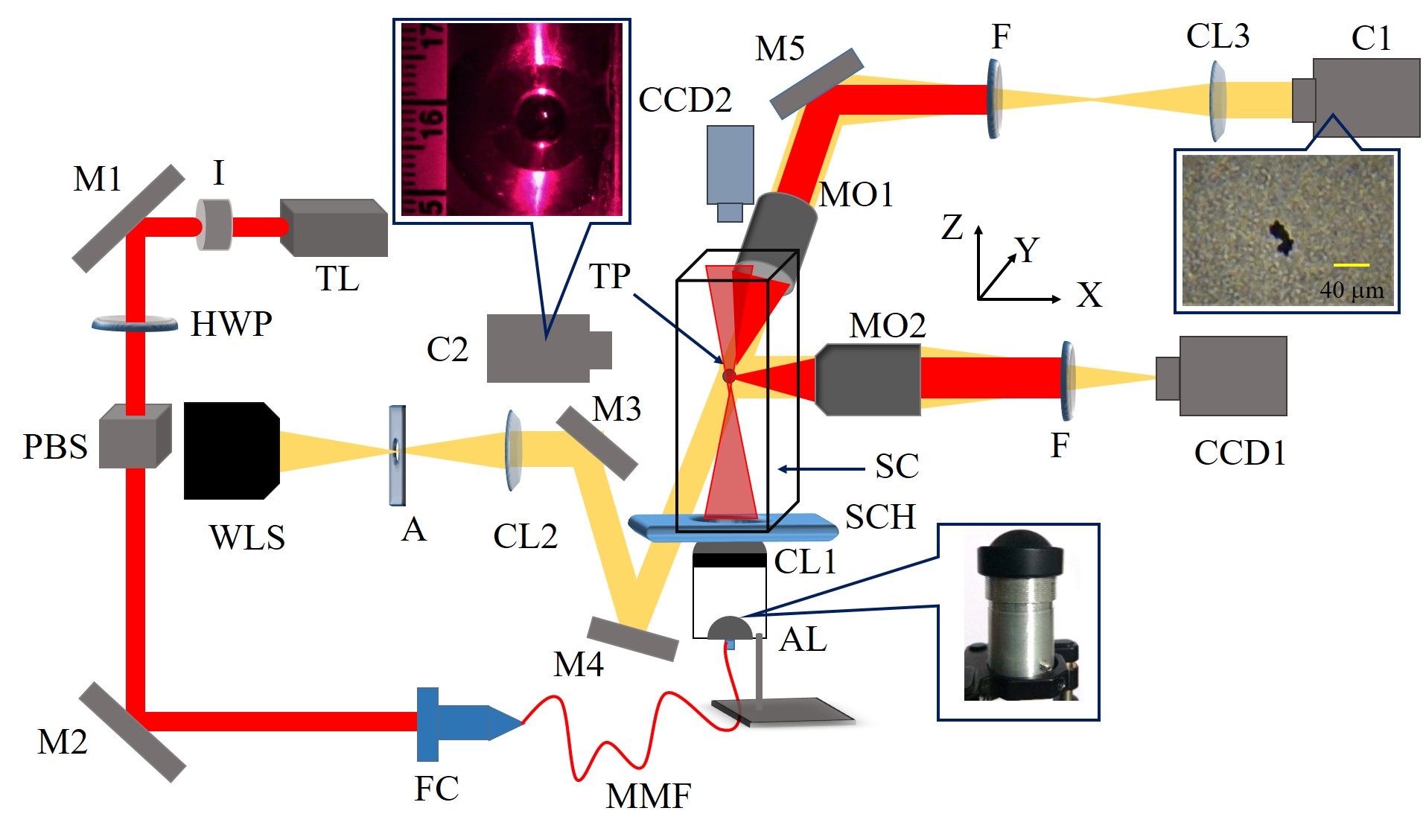}
	\caption{Schematic of the experiment. A: aperture; AL: Aspheric lens; C: Camera; CCD: Charge coupled device CL: Convex lens; F: Filter; FC: Fiber Coupler; HWP: Half wave plate; I: isolator; M: Mirror; MMF: Multimode fiber; MO: 10x objective; PBS: Polarizing beam splitter; SC: Sample chamber; TL: Trapping Laser; TP: Trapped particle; WLS: White light source} 
	\label{Fig2}
\end{figure}

We now describe the experiment towards measuring trapping efficiency for both beam profiles, keeping all other trapping parameters (i.e., laser power, beam size, etc.) invariant. A schematic of the experimental setup is shown in Fig.~\ref{Fig2} where we use a 671 nm laser source of maximum power 300 mW as a trapping beam for trapping printer toner particles that have very high absorptivity at our operating wavelength. Then, we couple the laser beam into a multi-mode fiber using the mirrors M1, M2, and the fiber coupler (FC) [see Fig.~\ref{Fig2}] after passing through an optical isolator (for preventing feedback from the fiber which destabilizes laser output) and a combination of a half-wave plate (HWP) and a polarizing beamsplitter (BS) [see Fig.~\ref{Fig2}] for changing the laser power in a controlled manner.  The output beam from the fiber is then collimated and focused into the sample chamber via a home-built mount which contains an aspheric lens (AL) for collimation and a 25 mm plano-convex lens (CL1) for focusing [see bottom right inset of Fig.~\ref{Fig2}]. This mount is then attached to a motorized translation stage TS, so that when the stage is translated, so is the trapping beam within the sample chamber. This is what employ for the drag force measurement.  

The trapped particles are imaged along in the $x$ and $y$ directions for determining the size and mass of the particle. For imaging, we use a white light source (WLS) which is collimated by CL2 and passes through the sample chamber and trapped particle with the help of mirrors M1 and M2. After that, the trapped particles are imaged on camera CCD1 in the $x$-direction using a $10x$ collection objective MO2, and a notch-filter F to block the trapping beam. In the $y$-direction, the particles are imaged on a Sony fast video camera C1 (1000 FPS) with the help of a 3-f imaging system, which is composed of a $10x$ objective lens (MO1), the lens CL3, and the lens placed inside the camera C1, where MO1 is used for the collection and another filter F is used to cut off the light at 671 nm. This 3-f imaging system provides a high contrast zoomed-in image of the trapped particles -  a representative image of which is shown in the top right inset of Fig.~\ref{Fig2}. In addition, another video camera, C2, is used to determine the axial position of a trapped particle in the $z$-direction by taking an image of the trapped particle along with a measuring scale affixed to the sample chamber, and imaging the motion of the trapped particle in order to measure the radial velocity. A representative zoomed-in image of the sample chamber with two particles trapped taken using C2 is shown in the top-left inset of Fig.~\ref{Fig2}.

\section{Results and Discussions}
\subsection{Comparison of Trapping force for Single and Multi-mode fiber}
We keep the laser power at 70 mW throughout the experiments and trap 20 particles using both the multi-mode and single-mode fiber, and make a comparison of the trap parameters for radial trapping between them. The results are shown in Table~\ref{tab:Radial Velocity and Threshold power}, with the number in parenthesis-denoting 1 $\sigma$ errors in the mean. First, we determine the average particle size (a) using the methodology described in Ref\cite{sil2020study}, where we observe that the average particle size for both beam profiles is almost the same [see table~\ref{tab:Radial Velocity and Threshold power}]. Then we determine radial trapping force by the viscous drag method, where we accelerate the stage TS with an acceleration of 0.1 $mm/s^2$ and thereby reach a maximum velocity of 5 mm/s along radially - so that the trapped particle also translates radially with the same acceleration. Hence, the drag force experienced by the particle increases till when it overcomes the trapping force, at which time the particle leaves the trap. We record the particle movement using our camera CCD1 while moving the stage, and perform a frame-by-frame analysis to measure the distance traversed by the particle before it leaves the trap. We  use the ImageJ software and correspondingly measure the velocity of the particle ($v_e$) at the point of escape from Newton's equations of motion. Thus, the escape velocity ($v_e$) achieved by the particle in the single-mode fiber trap is measured to be 0.33 (6) $mm/s$, while that in the multi-mode fiber trap is 2.53 (16) $mm/s$ [see Table~\ref{tab:Radial Velocity and Threshold power}]. We are then able to calculate the radial trapping force $F_{trap}$ by using the equation $F = 6 \pi \eta a v_e $ for both trap systems, assuming the particle to be spherical, where $\eta$ is the viscosity of the air, and $a$ is the trapped particle radius. The trapping force measurements come out to be 1.01 (17) $mm/s$ and 7.87 (51) $mm/s$ for single-mode and multi-mode fiber trap, respectively [see Table~\ref{tab:Radial Velocity and Threshold power}]. Thus, from the results, it is clear that the trapping force in the case of a multi-mode fiber trap is around eight times higher compared to that by a single-mode fiber trap.

\begin{table}[htbp]
	\centering
	\caption{\bf Comparison of trap parameters for Multimode and Gaussian beams}
	\begin{tabular}{cccc}
		\hline
		Trap parameters & Multi-mode (1)  & Single-mode (2) & Ratio (1)/(2)  \\ 
		\hline
		Average particle size ($\mu m$)  & 8.96 (0.30) & 8.15 (0.33)  & 1.10 (0.08)   \\
		Average $v_{e}$ (mm/s)  & 2.53 (0.16) & 0.33 (0.06)  & 7.67 (1.95)   \\
		Average $F_{trap}$ (pN) & 7.87 (0.51) & 1.01 (0.17) & 7.79 (1.87)  \\
		Average threshold power (mW) & 10(2) & 47(2)  & 0.2 (0.05) \\
		\hline
	\end{tabular}
	\label{tab:Radial Velocity and Threshold power}
\end{table}

Next, we measure the threshold power for trapping where we trap a particle at a moderate laser power and then reduce the laser power. While lowering, the trapped particle moves closer to the focus, where the intensity is high enough to provide enough photophoretic force to balance gravity. But, if we keep lowering the laser power, a point comes where the laser intensity is no longer able to generate a photophoretic force that can balance the particle's weight - thus, the laser power at which the particle leaves the trap is called the threshold power. The result is shown in Table~\ref{tab:Radial Velocity and Threshold power}, where we observe that the threshold power is around 47 (2) mW for single-mode trap and 10 (2) mW for Multi-mode trap. This signifies that the multi-mode beam profile can be trap particles at around 4.7 times $\%$ less power than the Gaussian beam, which indicates the multi-mode trap is about 4.7 times more stable than the single-mode trap. Hence, it is clear from our measurements that a multi-mode trapping beam is considerably more effective in trapping absorbing particles compared to a fundamental Gaussian beam. 

\subsection{Trapping force for different modes of Multi-mode beam profile}
\begin{figure}[h]
	\centering
	%\captionsetup{justification=centerlast}
	\includegraphics[scale=0.47]{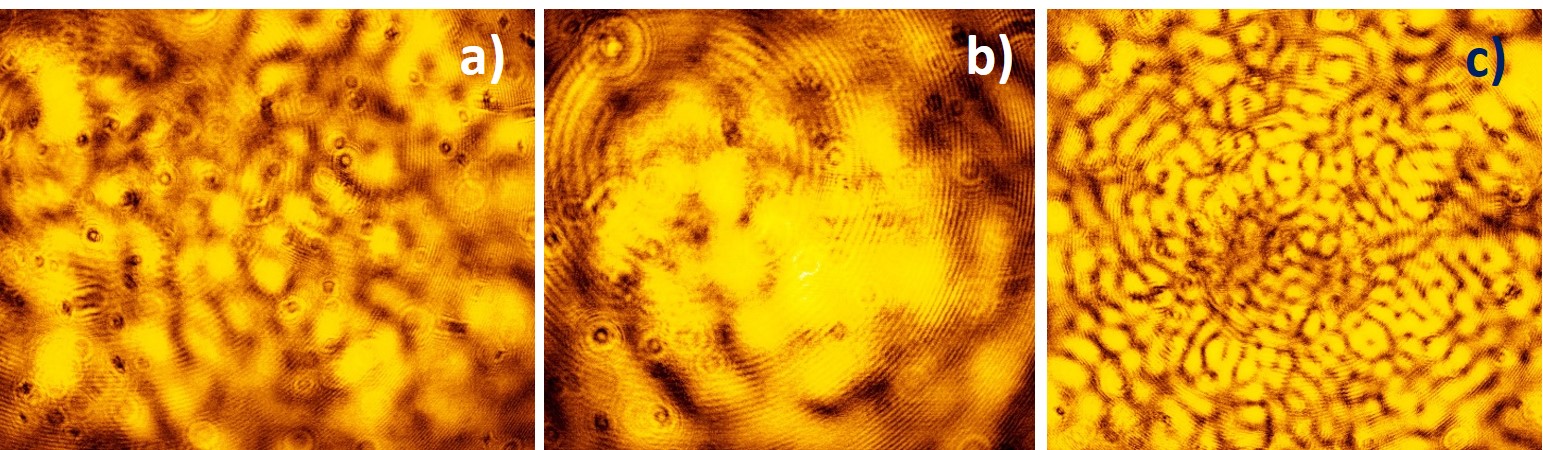}
	\caption{Speckle patterns created from Multimode fiber; a) Mode 1, b) Mode 2 c) Mode 3} 
	\label{Fig3}
\end{figure}

However, while performing the experiment for measuring the trapping force using the drag force method, we observe that the escape velocity of the trapped particle changes when we modify the speckle pattern, which signifies that not only the transverse extent and off-axis intensity would affect the trap efficiency, but also the speckle presents in the pattern. Thus, we systematically generate three different speckle patterns at the output of the multi-mode fiber by changing the coupling angle of the fiber coupler - so that different modes are excited inside the fiber, and their interference creates different types of final mode or speckle pattern at the output of the fiber. Three modes, which are hereafter referred to as Mode 1, Mode 2, and Mode 3, are shown in Fig.~\ref{Fig3} (a), (b) and (c), respectively.

In the experiment, we trap around 15 particles for each mode, and take the images of each by the cameras CCD1 (x-axis), C1 (y-axis), and C2 (axial position) [see Fig.~\ref{Fig2}]. We introduce another camera, CCD2, for monitoring speckle patterns for further analysis. Here, we also use the viscous drag method (discussed earlier) for radial trapping force measurement by measuring the radial escape velocity of a particle trapped using each mode. The average escape velocity ($v_e$) and the corresponding average radial trapping force ($F_{trap}$) are shown in Table~\ref{tab:Radial Velocity for different modes} for different modes, with the number in parenthesis-denoting 1 $\sigma$ errors in the mean as before. Note that, while doing the experiment we keep the laser power beam size invariant for each mode, and for determining the $F_{trap}$, we consider an experimentally measured average particle radius of 8.01(10) $\mu m$ and viscosity of air $\eta = \ 1.96 \times 10^{-5} \ kg/(m s)$ 

\begin{table}[htbp]
	\centering
	\caption{\bf Radial trapping force for all three mode}
	\begin{tabular}{ccc}
		\hline
		Mode name & Average $v_{escape}$ (mm/s)  &Average $F_{trap}$ (pN)   \\ 
		\hline
		Mode 1  & 2.64 (14) & 7.81 (45)    \\
		Mode 2  & 1.54 (12) & 4.57 (34)     \\
		Mode 3  & 1.93 (12) & 5.70 (35)  \\
		\hline
	\end{tabular}
	\label{tab:Radial Velocity for different modes}
\end{table}  

From Table~\ref{tab:Radial Velocity for different modes}, it is clear that the Mode 1 pattern provides 41\% and 27\% higher trapping force compared to Mode 2 and Mode 3, respectively. Also, Mode 3 provides gives 20\% higher trapping force than Mode 2. Thus, we can conclude that the speckle distribution and size definitely affect the efficiency of photophoretic trapping. To obtain a more quantitative understanding of this observation, we further measure the average speckle size for all three modes and determine the average intensity per speckle by counting the number of bright spots present in each mode pattern. We describe this in the next section.

\subsection{Numerical Simulation and Analysis}
\subsubsection{Speckle Size Measurement}
We know that speckle is a random distribution of light field - consisting of a multitude of dark and bright spots resulting from destructive and constructive interference\cite{goodman1976some}. There are different speckle parameters such as mean speckle size, contrast, intensity and polarization etc\cite{piederriere2004scattering}. But here, we only consider the mean speckle size, defined as the average size of bright or dark spots present in the pattern\cite{goodman1976some}. Thus, in order to find out the mean speckle size, we need to measure the Wiener spectrum of the pattern, which is the average intensities of all possible spatial frequency components of the pattern\cite{goodman1976some}. This can be done by calculating the normalized autocovariance function of the intensity speckle pattern obtained in the observation or image plane (x,y). Further, this function can be considered as the normalized autocorrelation function of the intensity, which has a zero base, and its width provides a good measurement of the average width of a speckle\cite{piederriere2004scattering,hamarova2014methods}. The methodology for finding out the speckle size is discussed in the Appendix Section 1. Thus, we determine the normalized auto-correlation intensity distribution ($C_I(i,j)$) of any given pattern image using the algorithm described by Eq.~ 5 in the Appendix. 

\begin{figure}[h]
	\centering
	%\captionsetup{justification=centerlast}
	\includegraphics[scale=0.52]{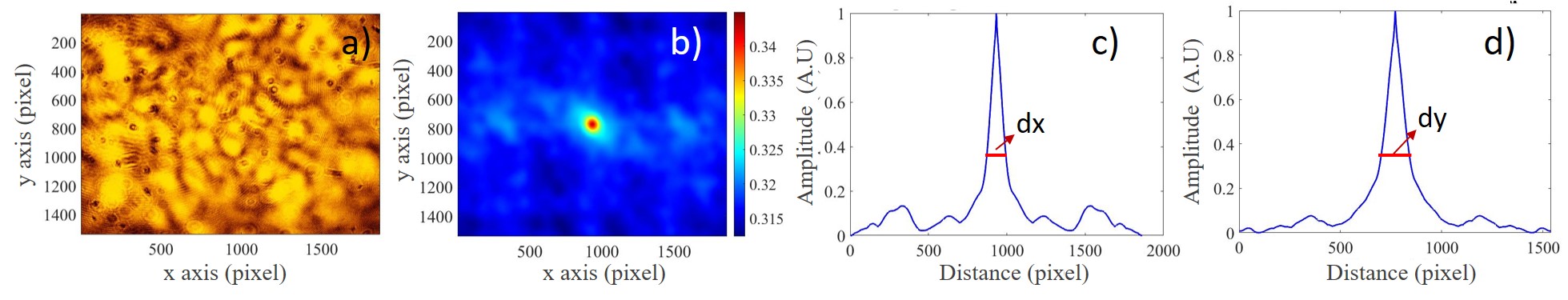}
	\caption{(a) Speckle pattern image of mode 1; (b) Normalized auto-correlation intensity distribution $C_I(i,j)$ profile of the image (a); c) \& (d) Horizontal profile $C_I(i,0)$ and  Vertical profile $C_I(0,j)$ of normalized auto-correlation function $C_I(i,j)$ (b), respectively}
	\label{Fig4}
\end{figure} 

However, for representation, the speckle pattern of Mode 1, and the corresponding normalized auto-correlation function $C_I(i,j)$ of that pattern are shown in Fig~\ref{Fig4} (a) and (b), respectively. Now, the mean speckle size is defined as a value where the horizontal (X) or vertical (Y) profile of normalized auto-correlation of intensity function $C_I(i,j)$ decays to 1/e \cite{hamarova2014methods}. So, $C_I(i,0)$ and $C_I(0,j)$ give the horizontal (X) and vertical (Y) profile of $C_I(i,j)$, which are shown in Fig. \ref{Fig4} (c) and (d), respectively. Then, we obtain the widths $dx$ and $dy$, where $C_I(0,dy) = C_I(0,dy) = 1/e$, for the horizontal ($x$) and vertical ($y$) directions, respectively [see Fig. \ref{Fig4} (c) and (d)].  

We observe experimentally that on increasing the size of the beam, the speckle size also increases correspondingly. Hence, if we consider the ratio between the speckle size and beam waist size at a particular plane of the respective pattern, the ratio should be invariant irrespective of the beam size, which implies that we can exactly determine the speckle size at any transverse plane along the laser propagation direction, if we know the beam size in that plane. We now describe the methodology of determining this ratio. Fig.~\ref{Fig4} (a) depicts the speckle image of Mode 1 of dimension (1540 $\times$ 1864) (pixel)$^2$  - implying the total length along the horizontal direction ($L_x$) and vertical direction ($L_y$), are 1864 and 1540 pixel, respectively [see Fig.~\ref{Fig4} (c) and (d)]. Next, we find the width $dx$ and $dy$ both horizontally and vertically to be 114 and 126 pixel, respectively, so the ratio along both the horizontal and vertical direction becomes $R_x = dx/L_x$ and $R_y = dy/L_y$, and finally take the average (R) between Rx and Ry. This algorithm is applied for the other two modes (Mode 2 and Mode 3). The speckle size along both $x-$ and $y-$axes, and the average speckle size for all three modes are shown in Table \ref{tab:Speckle size ratio}.

\begin{table}[htbp]
	\centering
	\caption{\bf Speckle size ratio for all three modes}
	\begin{tabular}{cccc}
		\hline
		Mode name & Horizontal (X) speckle  & Vertical (Y) speckle & Average speckle    \\ 
		& size ratio ($R_x$) & size ratio ($R_y$) & size ratio (R)  \\
		\hline
		Mode 1  & 0.061 & 0.082 & 0.071    \\
		Mode 2  &  0.226 & 0.184 & 0.205   \\
		Mode 3  & 0.026 & 0.029  & 0.027    \\
		\hline
	\end{tabular}
	\label{tab:Speckle size ratio}
\end{table}

Thus, using these ratios, we can find out the exact speckle size at the trapping region of the respective modes from a knowledge of the beam size at that region. However, the particles are trapped at a different position axially for each mode, so first we find out the exact $z$ position of the trapped particles by analyzing the camera images of C2 [see Fig.~\ref{Fig2}]. The mean $z$ positions of the trapped particles for each mode are shown in the first column of Table \ref{tab:Particle and speckle size}. Next, we find out the beam sizes at those $z$ positions by measuring the beam radii using the well-known knife-edge technique\cite{de2009measurement}, which are shown in the second column of Table \ref{tab:Particle and speckle size}. Then, we find out the speckle size at the respective z positions by multiplying the mean beam size (Table \ref{tab:Particle and speckle size} second column) with the respective average speckle size ratio (R) [Table \ref{tab:Speckle size ratio}] of each mode which are shown in the third column of Table \ref{tab:Particle and speckle size}. Finally, we determine the trapped particles' size and mass from their images, taken using cameras CCD1 and C1 [see Fig.~\ref{Fig2}] using the methodology given in Ref.~\cite{sil2020study}, and find out the average trapped particle diameter for each mode which we display in the last column of Table \ref{tab:Particle and speckle size}. This shows that the average diameter of trapped particles is almost the same for each mode.

\begin{table}[htbp]
	\centering
	\caption{\bf Mean speckle size and  particle diameter for all three modes}
	\begin{tabular}{ccccc}
		\hline
		Mode name & Mean z position  & Mean Beam size & Mean speckle size & Mean particle diameter    \\ 
		& (mm) & ($\mu m$) & ($\mu m$) & ($\mu m$)   \\
		\hline
		Mode 1  & 2.0 (0.1) & 235.97 (10.78) & 16.75 (0.77) & 16.02 (1.32) \\
		Mode 2  & 1.80 (0.16) & 218.17 (15.54) & 44.72 (3.19) & 14.76 (0.86)   \\
		Mode 3  & 1.90 (0.09) & 220.49 (10.01) & 5.95 (0.27) & 14.10 (0.64)  \\
		\hline
	\end{tabular}
	\label{tab:Particle and speckle size}
\end{table} 

Thus, we generate three modes pattern with different speckle sizes, which are 16.75 (0.77) $\mu m$, 44.72 (3.19) $\mu m$, 5.95 (0.27) $\mu m$ for Mode 1, Mode 2 and Mode 3, respectively. The radial trapping force is different for different modes, but interestingly for Mode 1, we get maximum trapping force where the mean speckle size and particle diameter are almost the same (see Table \ref{tab:Radial Velocity for different modes}). Thus, we may reasonably conclude that the trapping efficiency is better when the particle dimension and speckle size are comparable. Besides, we also observe that for both bigger and smaller speckle sizes compared to the particle size, the trapping force decreases. We attempt to understand this more elaborately in the next section.

\subsubsection{Average Intensity per speckle:} 
We make use of the fact that the photophoretic forces depends on laser intensity\cite{sil_2017,rohatschek_1995}, so that the average intensity per speckle will serve as a crucial parameter for controlling the trapping force. We therefore proceed to estimate the average intensity per speckle for each mode by counting the total number of bright spots present in the pattern, and determining their average intensity. This can be considered as an average intensity per speckle ($\langle I_{speckle} \rangle$), which notation we use hereafter. We first describe the methodology for counting the total number of bright spots present in each speckle pattern. 

\begin{figure}[h]
	\centering
	%\captionsetup{justification=centerlast}
	\includegraphics[scale=0.48]{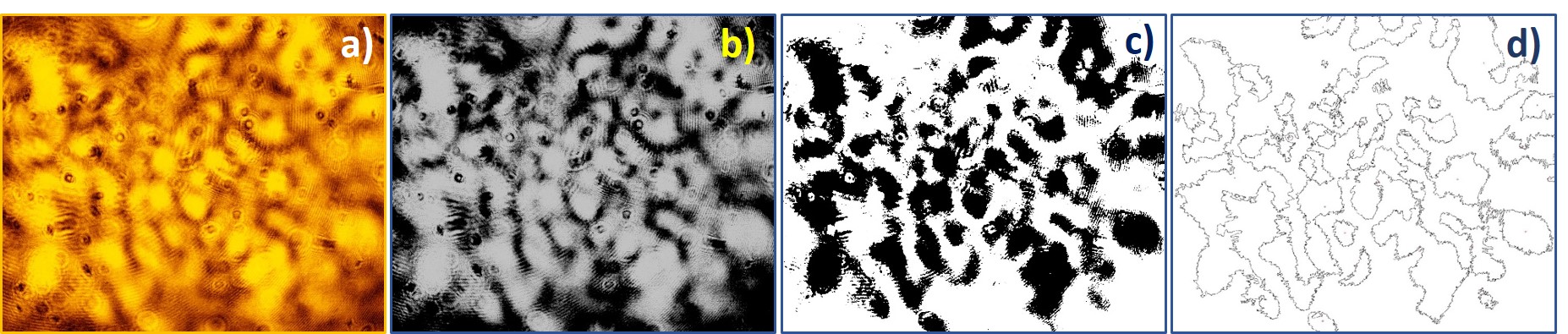}
	\caption{(a) Speckle image for Mode 1; (b) Green channel image of a); (c) After doing threshold of the image b); (d) Locate and count the Bright spot present in the pattern.}
	\label{Fig5}
\end{figure}

Thus, we consider an rgb speckle image of any mode, say Mode 1, which is shown in Fig.~\ref{Fig5} (a). We then split the image into three channels (red, blue, and green) using the ImageJ software, and work with the green channel image as it has good contrast, as shown in Fig.~\ref{Fig5}(b). We proceed to performing the threshold of that green channel image, i.e. the binary image as shown in Fig.~\ref{Fig5}(c). Note that we adjust the threshold value in such a way that, the bright spots in the green channel image [Fig.~\ref{Fig5}(b)] are converted into complimentary dark spots in the threshold image [Fig.~\ref{Fig5}(c)]. After that, we locate these dark spots with a curve using the software (`Analyze Particles' tool) by setting up the appropriate size ranges in pixels based on the speckle size. Finally, we use the software to count the total number of dark spots in the bounded region - which gives us a count of the high intensity speckles present in the pattern. 

For Mode 1 Mode 2 and Mode 3 [Fig.\ref{Fig3} (b) and (c)], we obtain numbers of 35, 10, and 308, bright spots respectively [see Table \ref{tab:Average Intensity per speckle}, second row]. Note that these number of bright spots should be invariant for any beam size, which we experimentally verify. Understandably, there also exists an inverse relationship between the speckle size and the number of bright spots present in the speckle pattern. On another note, the particles are trapped at different locations, which implies different beam sizes and speckle size as well for the respective modes [see Table \ref{tab:Particle and speckle size} first, second and third column]. Thus, we find out the mean speckle size which are depicted in the first row of Table \ref{tab:Average Intensity per speckle}. The laser power ($P$) is kept constant throughout the experiment for each mode. Hence, the laser power per speckle ($p$) for each mode should be $p = \frac{P}{N}$. Then we find out $\langle I_{speckle} \rangle = \frac{p}{A}$  where $A$ is the area of the average speckle size of the respective mode. The $p$ and $\langle I_{speckle} \rangle$  values for each mode are shown in Table \ref{tab:Average Intensity per speckle}.  
\begin{table}[htbp]
	\centering
	\caption{\bf Number of bright speckle and average intensity per speckle for the three modes}
	\begin{tabular}{cccc}
		\hline
		Mode name & Mode 1   &  Mode 2  & Mode 3   \\
		\hline
		Average speckle size ($\mu m$)   & 16.75 (0.77) & 44.72 (3.19)  & 5.95 (0.27)  \\
		Total number of bright speckle (N) & 35 & 10  & 308  \\
		Laser Power/speckle (p) (mW)   & 2.0 & 7.0  & 0.23  \\
		Laser Intensity/speckle $(\mu W/{\mu m}^2)$  & 7.13 (0.65)  & 3.50 (0.51)   & 6.50 (0.60)   \\
		Effective Intensity for Gaussian $\langle I_{eff} \rangle$ $(\mu W/{\mu m}^2)$ &   &1.62    & \\
		\hline
	\end{tabular}
	\label{tab:Average Intensity per speckle}
\end{table}

As shown in the Table \ref{tab:Average Intensity per speckle}, average intensity per speckle  is maximum for Mode 1 with $\langle I_{speckle} \rangle = 7.13 (0.65)~ \mu W/{\mu m}^2$ followed by Mode 3, which is 6.50 (0.60) $\mu W/{\mu m}^2$,  with Mode 2 being the lowest at 3.50 (0.51) $\mu W/{\mu m}^2$. Importantly, we observe a similar trend for the radial trapping force as shown in Table \ref{tab:Radial Velocity for different modes}, where the radial trapping force is maximum for Mode 1, followed by Mode 3 and Mode 2. Again, for the Gaussian beam, we determine the effective intensity, which is the actual intensity perceived by the particle. Note that the particle is smaller than the beam waist size, and thus does not perceive the entire beam intensity. This effective intensity is then given by an average of the intensity values of different non-overlapping sections of the beam where the particles of average diameter 16 $\mu m$ can be trapped\cite{sil2022trapping}. Besides, we observe from experiments that for the Gaussian beam, the average beam size where the particles are trapped is 200 $\mu m$. Thus, for the 200 $\mu m$ beam diameter, $\langle I_{eff} \rangle$ becomes 1.62 $\mu W/{\mu m}^2$. Note that, since Mode 1 gives the maximum trapping force experimentally, we use this mode to compare with the Gaussian beam. In our experiments, we compare the trapping force for both Multi-mode and Gaussian beam profiles by trapping the particles at the same beam size. So, the average intensity per speckle ($\langle I_{speckle} \rangle$) for 200 $\mu m$ beam size is 9.92 $\mu W/{\mu m}^2$. Hence, the experimentally measured force enhancement factor of eight also compares well with our numerical estimation of six.

\subsubsection{COMSOL simulation to determine temperature distribution across a trapped spherical particle due to laser heating}
\begin{figure}[h]
	\centering
	%\captionsetup{justification=centerlast}
	\includegraphics[scale=0.60]{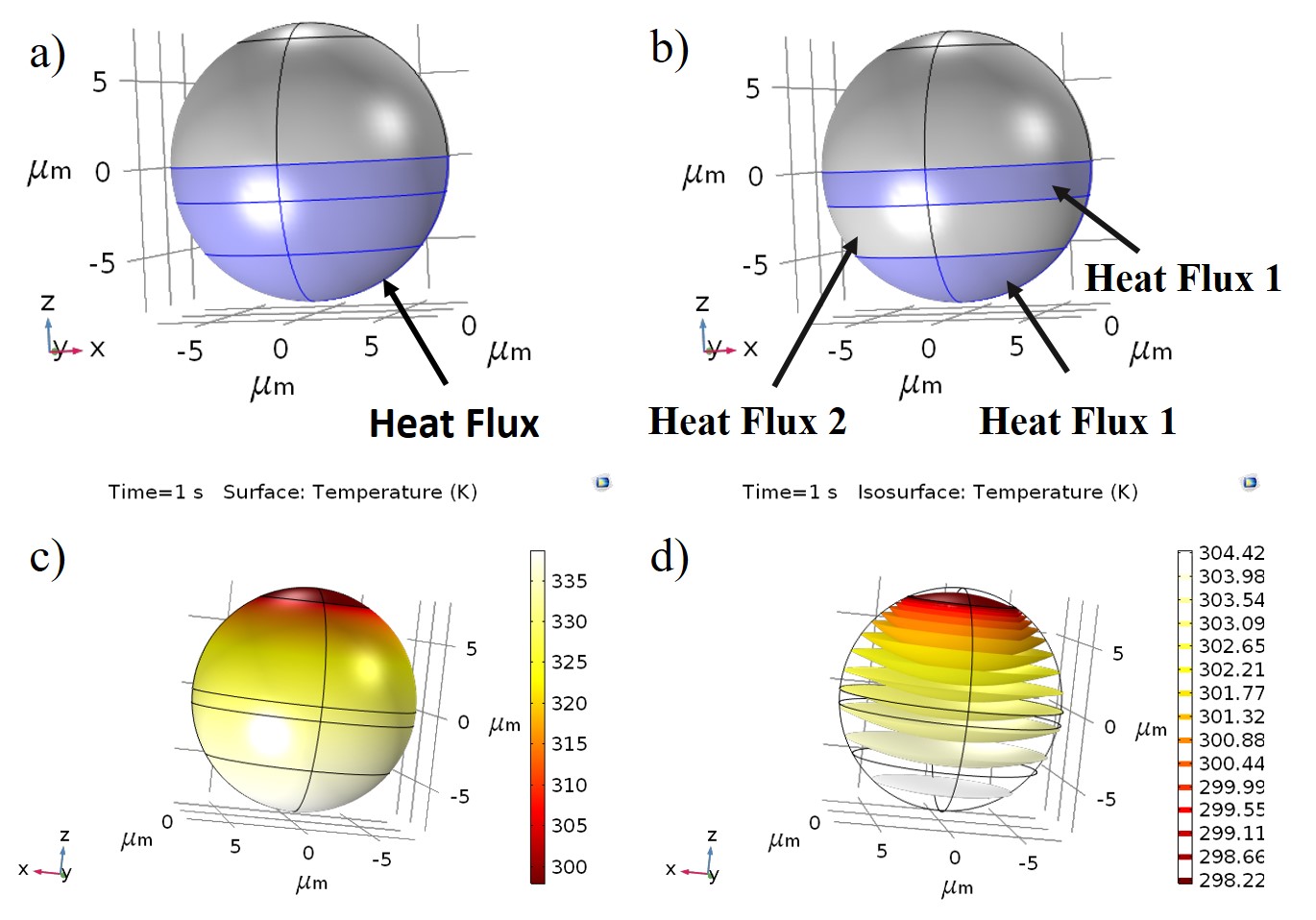}
	\caption{ Schematic for providing heat flux ($H$) to the lower surface of the particle (a) for Mode 1, Mode 2 of the  multi-mode fiber, and the Gaussian beam, (b) for mode 3. (c) Temperature distribution across the particle surface due to the laser intensity corresponding to the Mode 1 of the multi-mode beam profile; (d) Iso-surface of the temperature of the particle for input Gaussian beam.}
	\label{Fig6}
\end{figure}

We now numerically estimate the values of the photophoretic forces and radiation pressure force experienced by the particle. Hence, we employ the analytical formula for photophoretic $\Delta T$ and $\Delta \alpha$ forces acting on a spherical particle provided by Rohatschek in his semi-empirical model of photophoretic forces \cite{rohatschek_1995}, which we have described in detail in the Appendix (Section II). Note that the quantitative analysis of photophoretic forces acting on a particle is quite complex, as many factors are involved with this force - such as pressure, different parameters of light (beam profile, intensity, wavelength of the laser, etc.), and most importantly, particle properties (i.e., particle size, morphology, thermal conductivity, absorptivity, etc.)\cite{horvath_2014}. Thus, only semi empirical estimates of the forces are available from the literature. However, the photophoretic forces significantly depend on the temperature distribution across the particle's surface. Hence, we perform a COMSOL simulation to find out the temperature distribution across a particle due to laser heating. We use the 'Heat transfer in solid (time-dependent) model' and assume a spherical particle of radius 8 $\mu m$ again, as our experiments revealed this to be the average radius of the trapped particles. Further, we choose the incident heat flux (H) as $H = \chi <I>$, where $\chi$ is the absorptivity of the particle, and $<I>$ is the average laser intensity which can be ${\langle I_{speckle} \rangle}$ for a multi-mode and ${\langle I_{eff} \rangle}$ for a Gaussian beam. Note that this heat flux is considered as a spatially distributed heat source on the particle surface - introduced at the lower hemisphere [see Fig.~\ref{Fig6}(a) and (b)]. 

Now, according to our earlier estimation of the speckle size, it is clear that for Mode 1 (16.75 $\mu m$), the speckle diameter is comparable to our particle diameter (16 $\mu m$), while both for Mode 2 and the Gaussian beam, the speckle (44.72 $\mu m$) and waist diameter ($\sim~200 ~\mu m$), respectively, are much bigger than the particle diameter. Therefore, in these cases, we fill up the lower hemisphere of the particle by the laser beam - to replicate which, we set the heat flux (H) over the entire region of the lower hemisphere of the particle as shown in Fig.~\ref{Fig6}(a).
The situation is more complex for Mode 3. Here, the average speckle size is 5.95 $\mu m$, so that the lower hemisphere experiences alternate bright and dark regions of the trapping light, both radially and axially.  Thus, to simulate this situation in the model, we create two partitions on the lower hemisphere of the particle - one at -5 $\mu m$,  and the other at -2 $\mu m$ from the bottom of the lower hemisphere [refer to the -8 $\mu m$ position in Fig.~\ref{Fig6}(b)]. Hence, we obtain three domains, of which the lowest one is of diameter 6 $\mu m$, so that we set this as `Heat flux 1' where $H = \chi * {\langle I_{speckle} \rangle}_{Mode \ 3}$. Next, we have the middle region, once more of diameter 6 $\mu m$, which we set `as Heat flux 2'. Note that here the intensity is virtually zero (since we assume a volume speckle field, where the bright and dark speckles are distributed uniformly in all three dimensions) , as the particle encounters a dark region here [see Fig.~\ref{Fig6}(b)]. Finally, for the uppermost region of 4 $\mu m$ diameter at the lower hemisphere, we again set `Heat flux 1' [see Fig.~\ref{Fig6}(b)]. Moreover, as the particle blocks the beam, we do not provide any heat flux at the upper hemisphere and set the ambient temperature of the particle at 298 K. The thermal conductivity, density, and specific heat of the printer toner particle we trap are provided as user-defined values.  

With this arrangement for multi-mode and Gaussian beams, we carry out our simulations. The temperature distribution across the particle surface are noted down for all particular simulations. A representative temperature distribution across the particle surface for Mode 1 of the multi-mode profile is shown in Fig.~ \ref{Fig6}(c). Note that we calculate the $\Delta T_s$ - which determines both the $\Delta T$ and $\Delta \alpha$ forces - by calculating the temperature difference between two hemispheres, and for that, we take very thin iso-surface temperature shells of both upper and lower hemispheres of the particle and correspondingly measure the average values of the lower hemisphere ($T_1$) and upper hemisphere ($T_2$) which is shown in Fig.~\ref{Fig6} (d). The difference between $T_1$ and $T_2$ gives the $\Delta T_s$ value, while we approximate the $\overline{T_s}$ by taking the average between $T_1$ and $T_2$ without going into too much complexity. The results for $T_1$, $T_2$, $\Delta T_s$ and $T_s$ for all speckle patterns of the multi-mode fibre are depicted in Table \ref{tab:Average Temp particle surface} 

\begin{table}[htbp]
	\centering
	\caption{\bf Temperature difference across the particle surface for all beam profiles}
	\begin{tabular}{ccccc}
		\hline
		Beam Profile & $T_1$ (K)  & $T_2$ (K) & $\Delta T_s$  & $\overline{T_s}$ \\ & At lower hemisphere & At upper hemisphere & ($T_1$ - $T_2$) (K)  & (K)  \\
		\hline
		Mode 1  & 325.31 (2.49) & 307.01 (0.82)  & 18.30 (1.67) & 316.16 (1.66)  \\
		Mode 2 & 311.41 (1.93) & 302.42 (0.63)  & 8.99 (1.31) & 306.92 (1.28)  \\
		Mode 3 & 313.91 (1.46) & 303.14 (0.47)  & 10.77 (0.99) & 308.53 (0.96) \\
		\hline
	\end{tabular}
	\label{tab:Average Temp particle surface}
\end{table}  

As the heat flux ($H$) is proportional to the laser intensity for the same particle, we obtain higher temperature difference at higher intensity, resulting in a maximum $\Delta T_s$ value for Mode 1 - by a factor of about 2.1 and 1.7 over Mode 2 and Mode 3, respectively. However, while the average intensity per speckle $\langle I_{speckle} \rangle$ for Mode 2 is 46\% less compared to Mode 3, the $\Delta T_s$ value is only 17\% less. This occurs since for Mode 3, some regions of the lower hemisphere of the particle interacts with the dark regions of the optical mode, which lowers the average temperature of that region, and thereby decreases the temperature difference between the two hemispheres. The results for  $T_1$, $T_2$, $\Delta T_s$ and $T_s$ for the Gaussian beam and Mode 1 of multi-mode beam for the same beam size of 200 $\mu m$ are shown in Table \ref{tab:Average Temp particle surface_Gaussian_MM} . We see from the table that the $\Delta T_s$ value for Mode 1 of the multi-mode profile is around six times higher compared to the Gaussian beam (generated from single-mode fiber), which is similar to the numerical estimation of intensities for both the beam profiles. 

\begin{table}[htbp]
	\centering
	\caption{\bf Temperature difference across the particle surface for mode 1 of multi-mode and
		Gaussian beam}
	\begin{tabular}{ccccc}
		\hline
		Beam Profile & $T_1$ (K)  & $T_2$ (K) & $\Delta T_s$  & $\overline{T_s}$ \\ & At lower hemisphere & At upper hemisphere & ($T_1$ - $T_2$) (K)  & (K)  \\
		\hline
		Gaussian  & 304.20 & 300.04  & 4.15 & 302.12  \\
		Mode 1 & 335.99 & 310.54  & 25.46 & 323.26  \\
		\hline
	\end{tabular}
	\label{tab:Average Temp particle surface_Gaussian_MM}
\end{table}  

\subsubsection{Calculation of total force acting on a spherical particle}
\begin{figure}[h]
	\centering
%	\captionsetup{justification=centerlast}
	\includegraphics[scale=0.55]{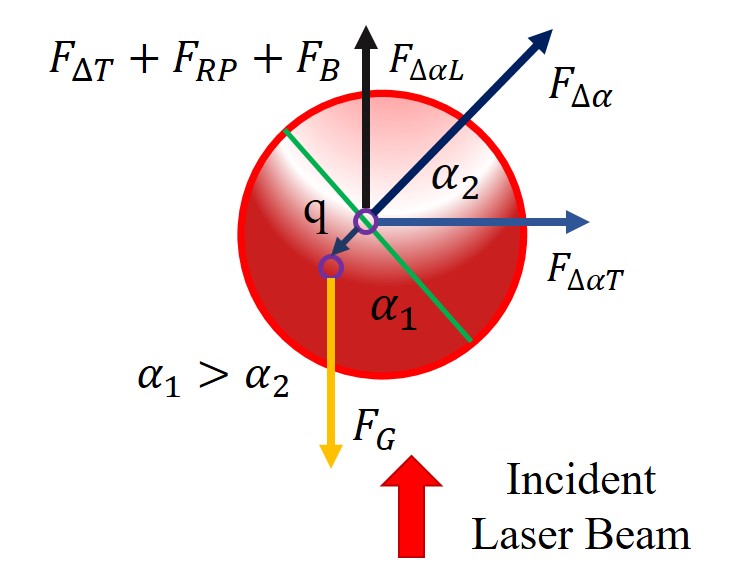}
	\caption{ Illustration of all the forces acting on a spherical particle}
	\label{Fig7}
\end{figure}

We now consider a force balance on the particle, by considering the effects of the photophoretic $\Delta T$ force, radiation pressure force $F_{RP}$ and buoyancy force $F_B$ - that always point along the laser propagation direction (vertically upward in our configuration) - and the gravity $F_G$, that points away from the laser (vertically downwards in our configuration) - as shown in Fig.~\ref{Fig7}. Further, there also exists the photophoretic $\Delta \alpha$ force, which is a body-fixed force, and so is directed from higher $\alpha$ to lower $\alpha$ as depicted by the green arrow in Fig.~\ref{Fig7}. All the estimated forces acting on the particle for all modes of the multi-mode fiber and the Gaussian beam are summarized in Table \ref{tab:Force calculation}. 

First, we calculate the photophoretic $\Delta T$ force by plugging in the $\Delta T_s$ values [see the fourth column of Table \ref{tab:Average Temp particle surface}] into the Eq. \ref{eq1},
\begin{equation}
	F = D  \ \frac{p*}{p} \ a \Delta T_s
	\label{eq1}  
\end{equation}
where, $D$ denotes a constant, determined entirely by the state of the gas and $p*$ is the characteristic pressure that depends on particle radius $a$, $p$ is the atmospheric pressure and $\Delta T_s$ is the temperature difference across the particle surface [For more details see Appendix Section 2]. The values of $F_{\Delta T_s}$ for all modes of the multi-mode fiber are shown in the second column of Table \ref{tab:Force calculation}. Next, we calculate the radiation pressure force using the formula $F_{RP} = \pi a^2 \frac{I}{c} (1 + R)$, where R is the reflectivity of the particle. For absorbing particles, this should be almost negligible, but we choose R = 0.1 as an upper limit. For intensity $I$, we take the average intensity per speckle ($\langle I_{speckle} \rangle$) for all  modes from the multi-mode fiber, and the effective intensity ($\langle I_{eff} \rangle$) for the Gaussian beam  [see Table \ref{tab:Average Intensity per speckle}], which are shown in the third column of Table \ref{tab:Force calculation}. Then, we calculate the gravitational force of the particle, and since we consider the particles to be spherical, $F_G = \frac{4}{3} \pi a^3 \rho = 30.88$ pN, where $a$ is the radius ($~\sim 8 \mu m$), and $\rho$ is the density of the particle. The buoyancy force, $F_B = \rho_{air} g V = 0.038$ pN, is negligible compared to the $F_G$, where $\rho_{air}$ is the density of air, $g$ is the gravitational constant, and $V$ is the volume of the particle. Finally, we calculate the photophoretic $\Delta \alpha$ force using the Eq. \ref{eq2} as depicted below (discussed in detail in Appendix),
\begin{equation}
	F_{\Delta \alpha} = \phi * B_1 =\frac{3}{4} D \frac{1}{\left(\frac{p}{p^*} + \frac{p^*}{p}\right)} a  (\overline{T_s} - T_i) \Delta \alpha 
	\label{eq2} 
\end{equation} 
The $F_{\Delta \alpha}$ values are shown in the sixth column of Table \ref{tab:Force calculation}, where $T_i$ is 298 K and the $\overline{T_s}$ values are taken from the last column of Table \ref{tab:Average Temp particle surface}. Note that it is impossible to know the exact distribution of accommodation coefficient $\alpha$ value over the particle surface, so that based on the literature \cite{rohatschek_1995}, we assume the $\alpha_1$ and $\alpha_2$ of the particle as 0.9 and 0.8 - giving $\Delta \alpha$ value to be 0.85. Now, while we know that $F_{\Delta \alpha}$ dominates over $F_{\Delta T}$ at atmospheric pressure, in our case it is the $F_{\Delta T}$ values which dominate over $F_{\Delta \alpha}$, as the thermal conductivity of the particles we use is minimal, (0.072 W/(m.K), which creates a substantial temperature difference across the particle surface.

\begin{table}[htbp]
	\centering
	\caption{\bf Calculation of all forces acting on the particle}
	\begin{tabular}{ccccccc}
		\hline
		Beam Profile & $F_{\Delta T_s}$ (pN) &$F_{RP}$ (pN) & $F_G$ (pN) & $F_B$ (pN)  & $F_{\Delta \alpha}$ (pN) & $F_{trap}$ (pN) \\
		\hline
		Mode 1   & 130.10 (11.87) & 4.56 (0.42)  & 30.88 & 0.038  & 9.68 (0.94) & 7.81 (0.45) \\
		Mode 2  & 63.91 (9.28) & 2.24 (0.32)  & 30.88 & 0.038  & 4.75 (0.68) & 4.57 (0.34) \\
		Mode 3  & 76.57 (7.04) & 4.16 (0.38)  & 30.88 & 0.038  & 5.61 (0.51) & 5.70 (0.36) \\
		\hline
	\end{tabular}
	\label{tab:Force calculation}
\end{table} 

Similarly, we calculate all the forces acting on the particle for both the Gaussian beam and mode 1 of multi-mode beam with the same beam size, which are shown in Table \ref{tab:Force calculation Gaussian_MM}. For the calculation of the photophoretic forces, we take the values of $\Delta T_s$ and $T_s$ values from Table \ref{tab:Average Temp particle surface_Gaussian_MM}.  

\begin{table}[htbp]
	\centering
	\caption{\bf Calculation of all forces acting on the particle}
	\begin{tabular}{ccccccc}
		\hline
		Beam Profile & $F_{\Delta T_s}$ (pN) &$F_{RP}$ (pN) & $F_G$ (pN) & $F_B$ (pN)  & $F_{\Delta \alpha}$ (pN) & $F_{trap}$ (pN) \\
		\hline
		Gaussian & 29.52 & 1.38  & 30.88 & 0.038  & 2.20 & 1.01  \\
		Mode 1   & 181.01 & 6.95  & 30.88 & 0.038  & 13.46 & 7.79 \\
		\hline
	\end{tabular}
	\label{tab:Force calculation Gaussian_MM}
\end{table} 

Since the $F_{\Delta \alpha}$ force is responsible for radial trapping (as we have mentioned earlier), the experimentally measured $F_{trap}$ values (shown in the last column of Table \ref{tab:Force calculation} and \ref{tab:Force calculation Gaussian_MM}) can be compared with the $F_{\Delta \alpha}$ values, obtained numerically. It is clear from Table \ref{tab:Force calculation} and \ref{tab:Force calculation Gaussian_MM} that we achieve good agreement between these values. It is also important to note that in the experiments, we measure only the radial component of the $F_{\Delta \alpha}$ force, which is not the case in the numerical estimation - so that it is reasonable to expect that the experimentally measured values would be lower than that estimated by the simulations. This is indeed what we obtain, as is clear from  Table \ref{tab:Force calculation} and \ref{tab:Force calculation Gaussian_MM}. Note that, for Mode 3, we obtain a lower numerical value than the experimental one, as the size of bright and dark spots present in the pattern is lower than the trapped particle size, which might affect the numerical estimation of the number of spots illuminating the particle.

Now, in our system, a particle is confined in a position axially when the gravitational force ($F_G$) balances the other three forces, $F_{\Delta T}, F_{RP}$ and $F_B$) [see Fig. \ref{Fig7}], though $F_{\Delta T}$ dominates the others. It can be observed from Table \ref{tab:Force calculation} and \ref{tab:Force calculation Gaussian_MM} that in general, for the multi-mode beam profile, the total upward force ($F_U = F_{\Delta T} + F_{RP} + F_B$) is significantly larger compared to the gravity $F_G$, as a result of which particles should shoot upwards in the propagation direction, and should thus not be confined. However, we do observe very strong and stable trapping in our experiments with the multi-mode fiber with the same beam parameters used in the simulation. We now attempt to explain this discrepancy using a simple model.  

\begin{figure}[h]
	\centering
	%\captionsetup{justification=centerlast}
	\includegraphics[scale=0.65]{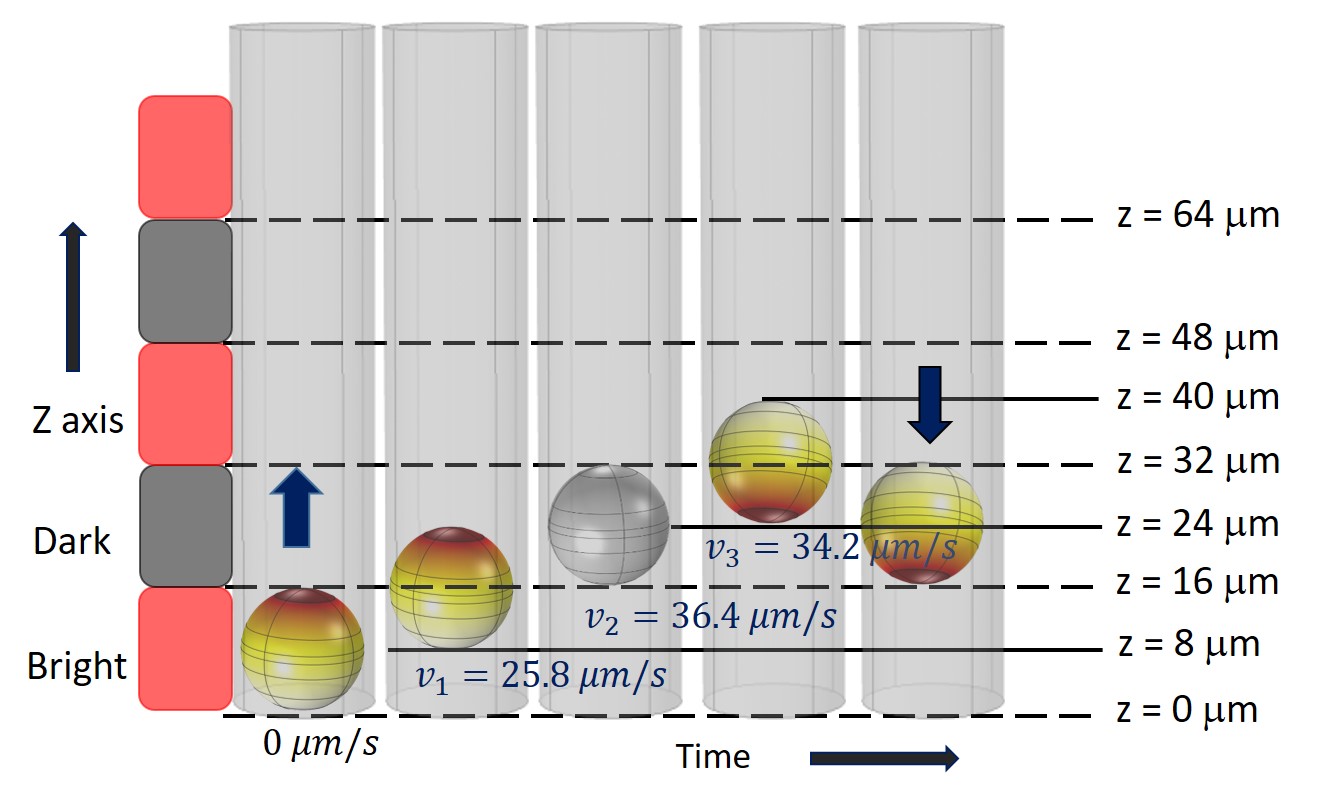}
	\caption{Model for particle oscillation along the z direction while being confined.}
	\label{Fig8}
\end{figure}

\begin{figure}[h]
	\centering
	%\captionsetup{justification=centerlast}
	\includegraphics[scale=0.50]{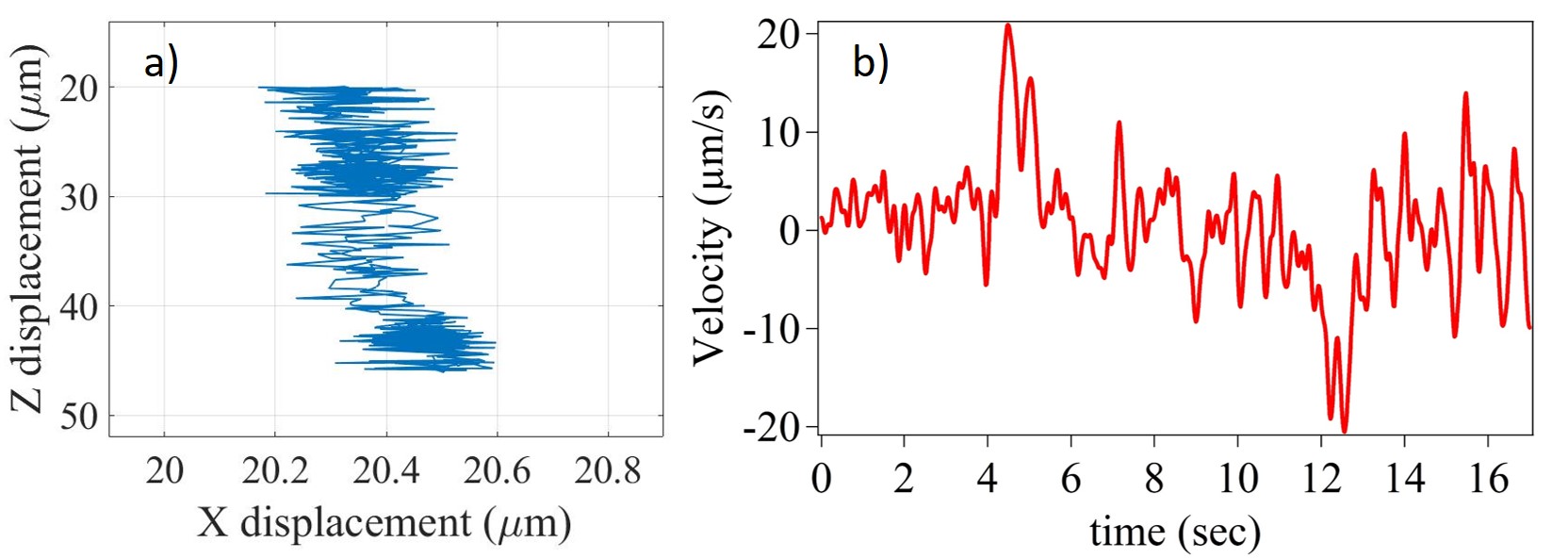}
	\caption{(a) X-Z plot of a trapped particle’s trajectory using multimode fiber of Mode 1 profile (b) Corresponding velocity plot of that trajectory along the z direction.}
	\label{Fig9}
\end{figure}

In order to simulate the multi-mode profile consisting of alternate bright and dark regions, we assume a beam structure in which these regions are stacked one after another axially, as shown in the Fig.\ref{Fig9}. For simplicity, we assume that the particle experiences bright and dark spots in sequence, which may be the case if there is a small angle between the beam axis and the trajectory of the particles. Further, when we perform the experiments to compare the trapping forces for the multi-mode and single-mode cases, the experimentally measured particle location data shows that the average beam waist diameter where the particles are trapped for the multi-mode profile is around 200 $\mu m$. Thus, in the simulation, we start from this position, i.e., assume z = 0 $\mu m$ here (depicted as a dotted line in Fig. \ref{Fig8}), and correspondingly measure average intensity per speckle ($\langle I_{speckle} \rangle$). Once again, we consider the speckle size for the beam profile corresponding to Mode 1, since this is the mode we choose for the experiments to compare performance. Now, using our earlier estimation of all forces acting on the particle, we define a resultant force ($F_{RE}$), which is [see Fig.~\ref{Fig8}]:
\begin{equation}
	F_{RE} = F_U - F_G
	\label{eq3} 
\end{equation}
where $F_U = F_{\Delta T} + F_{RP} + F_B$. It is clear that $F_U >> F_G$, so that a particle experiences a force axially and moves with a resultant acceleration $a_r = \frac{F_{re}}{m} - g$, where $m$ is the particle mass and $g$ is the gravitational constant.  

Here, we assume that initial velocity ($u$) at the starting point is zero (i.e., at z = 0 $\mu m$), and then determine the velocity $v_1$ when the particle moves $8 \mu m$ (h) using Newton's motion laws, viz. $v = \sqrt{u^2 + 2 a_r h}$, which gives $v_1 = 25.8~ \mu$ m/s [see Fig.~\ref{Fig8}]. After the particle traverses 8 $\mu m$ from the initial position axially, we recalculate all the forces (we ignore the viscous drag by the air for simplicity), and determine $F_{RE}$ using Eq.\ref{eq3}, followed by the resultant acceleration $a_r$ of the particle at the new axial position. Since, $F_{RE} >> 1$, the particle continues to move in the upward direction with an estimated velocity $v_2 = 36.4~ \mu$m/s [see Fig.~\ref{Fig8}]. After this, however, the particle is at z = 16 $\mu m$, and arrives in a dark region of the beam, where the photophoretic $\Delta T$ and $F_{RP}$ forces are almost zero. So, the resultant acceleration $a_r$ of the particle would be $-g$, but due to the initial large acceleration of the particle - it continues to move in the upward direction by another 8 $\mu m$ - albeit with a reduced velocity $v_3= 34.2~ \mu$m/s. Interestingly, at this position (i.e., at $z=24~\mu$m, the upper surface of the particle interacts with the bright region of the beam, while the lower region samples a dark region. As a result, the $F_{\Delta T}$ force and $F_R$ force are reversed, and directed towards gravity. Hence, the particle falls under gravity almost immediately, and back into a bright speckle again [see Fig.~\ref{Fig8}]. Thus, the particle undergoes a stable oscillation in the axial direction, and remains confined in the photophoretic trap, even with the laser intensity generating a photophoretic force higher than the gravitational force corresponding to the weight of the particle. 

A crucial issue now is to verify whether the mechanism we suggest for trapping from our simulations is also observed in experiments. This is indeed the case - and we observe clear signatures of trapped particles oscillating in the $z$ direction using Mode 1 [see Video1 in Appendix] of the multi-mode fiber. We have also quantified the average axial oscillation by tracking the trapped particle's position along the $x-z$ and $y-z$ planes by analyzing videos of its motion using the Matlab software. For representation, the $x-z$ trajectory of a trapped particle, and the corresponding velocity plot along the $z$ axis is shown in Fig.~\ref{Fig9}(a) and (b), respectively. We determine the average $z$ oscillation of particles to be 29.4(4.1) $\mu m$ from the data of 15 trapped particles of similar size. The oscillation amplitude is in reasonable agreement (around 17 $\%$) with the value provided by our simulations ($\sim 24~\mu$m). Also, the maximum velocity we measure is around 20 $\mu m/s$, which is about 40 $\%$ different from the simulations ($\sim 34.2~\mu$m/s), but this difference could well be due to the fact that we have ignored the drag force by air in the simulation, and also the fact that our camera has a limited frame rate [60 FPS in this case]. 

Very interestingly, we observe oscillations of around 0.4-0.5 $\mu$m in the radial direction as well in Fig.~\ref{Fig9}(a), which are of almost constant amplitude as the particle moves axially. This is understandable since the multi-mode beam has a speckle structure in all three dimensions, but the intensity of the speckles fall off faster in the radial direction (close to a Gaussian profile) compared to that in the axial direction - so that the particle displacement amplitude is smaller. 

\begin{figure}[h]
	\centering
	%\captionsetup{justification=centerlast}
	\includegraphics[scale=0.50]{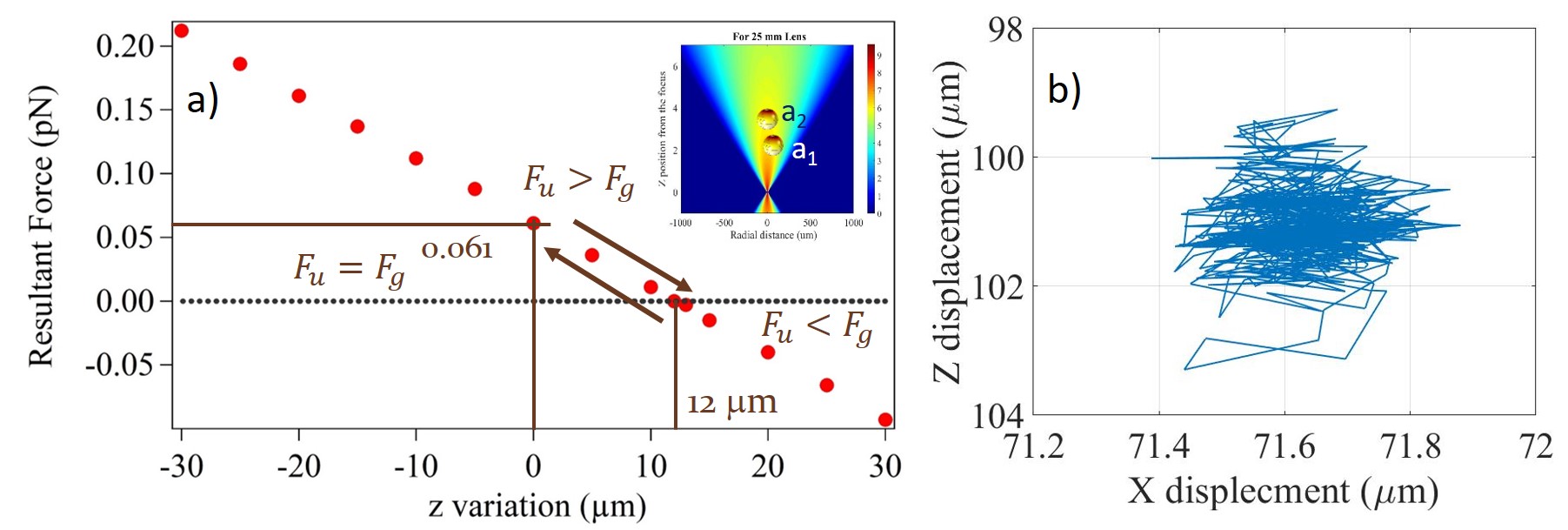}
	\caption{(a) Simulation for finding out the z drifting of a  trapped particle in case of Gaussian beam.(b) Experimental X-Z plot of a trapped particle’s trajectory using Gaussian beam}
	\label{Fig10}
\end{figure} 

Finally, we carry out a similar exercise for a particle trapped in a Gaussian beam. A calculation of all forces based on the location where the particle is trapped, as described in Table \ref{tab:Force calculation Gaussian_MM}, reveals that the total upward force $F_U$ is slightly higher than the downward force $F_G$. As a result, the particle can move in the upward direction. Let us assume that the location of the trapped particle is $a_1$ as shown in the right inset of Fig.~\ref{Fig10}(a), where we represent the propagation of Gaussian beam focused by a convex lens of focal length 25 mm. Since $F_U > F_G$ at this location, the particle moves along the upward direction. In the simulation, we move the particle by small step in this direction, and similar to the previous case, calculate all relevant forces.  So, when the particle moves by 12 $\mu m$  from the initial position as depicted by $a_2$ in the right inset of Fig.~\ref{Fig10} (a), we obtain $F_U = F_G$, so that the resultant force becomes zero. Ideally, the particle can be stably confined at the $a_2$ position, but due to perturbations such as laser intensity fluctuations, air turbulence, etc. the particle may well oscillate around this equilibrium position, which is shown as a black dotted line in Fig.~\ref{Fig10}(a). 

This is exactly what we observe in experiments. The average $z$ oscillations we measure in  particles trapped by a Gaussian beam [see Video2 in Appendix] is 9.6 (1) $\mu$m (from 15 sets of data), which is again in reasonable agreement (around 20 $\%$) from the simulation results. The $x-z$ trajectory plot of one of the trapped particles is shown in Fig.~\ref{Fig10}(b). Radial oscillations are also observed, but these are on the average between 0.2-0.3 $\mu$m, with a few oscillations reaching an extent of 0.4 $\mu$m. This is due to the fact that in comparison to a multi-mode beam profle - a Gaussian beam has more drastic intensity variation in the radial direction as compared to that in the axial direction (pure Gaussian versus quadratic). Thus, for both multi-mode and Gaussian beams, axial and radial oscillations are observed, and may contribute significantly to the stable trapping of particles. Multi-mode beam profiles, changing less rapidly in intensity both radially and axially compared to Gaussian ones for the same focusing lens, offer considerably more robust trapping due to the larger dynamic equilibrium range in both dimensions. 

A relevant question to ask here may be why a particle reaches an intensity region in the multi-mode profile which produces much higher photophoretic force than that required to balance its inertia. This, we believe, may to be due to the fact that we use a simple trapping chamber that is not sealed in any manner, so that particles falling are entirely exposed to microscopic air currents and turbulence, which may well exert instantaneous forces much larger than photophoretic forces.  Note that the large axial trajectories observed in the multi-mode case may also be due to the fact that the particle will continue its upward trajectory unless it comes into contact with a dark zone, at which point its trajectory will reverse. Thus, the spatial dynamic range of the particle oscillations are greater in the multi-mode fiber in almost all cases than that for single-mode fiber. We also believe that these oscillations are demonstrative of the existence of a restoring force in the case of photophoretic trapping even in the axial direction. The magnitude of the restoring force appears to be higher for a multi-mode fiber compared to a single-mode case, with the presence of dark zones contributing in the final dynamic equilibrium achieved by the particle. In addition, the threshold power of trapping in the case of multi-mode fibers (see Table \ref{tab:Radial Velocity and Threshold power}) is also lower than that for a pure Gaussian trap, since the intensity per speckle is much higher than the overall intensity measured from the beam waist size - something which is not the case for a Gaussian beam. 

\section{Conclusion}
In conclusion, we employ a single multi-mode fiber to develop a robust three-dimensional optical trap for trapping absorbing particles in air employing photophoretic forces. We observe that the intensity profile created by the multi-mode fiber provides around eight times higher trapping force compared to that produced by a single-mode fiber that produces a pure Gaussian beam. This is because the intensity a particle experiences in the speckle pattern generated by a multi-mode fiber is higher than the beam profile at the output of a single-mode fiber, so that the trapping force is correspondingly higher. Our studies reveal that a beam profile where the speckle size is similar to the particle size produces the strongest traps, using which we achieve axial and radial velocities of 5 $mm/s$, which is presently limited by our experimental capabilities.  We validate our experimental results by developing a COMSOL-based simulation to calculate all the forces experienced by a trapped particle, and applying force balance to study the particle dynamics. Our simulations reveal clear axial oscillations of the trapped particles in the case of both single and multi-mode fibers, with the multi-mode having higher spread due to the inherently high intensity of the individual speckles that constitute the beam profile. We validate our simulation results with experimental observations, where both single and multi-mode fibers give rise to axial as well as radial particle trajectories. The trajectories in the multi-mode case have considerably higher spread compared to the single mode case, as confirmed by our simulations. Indeed, our results also point out to the existence of an effective restoring force on the trapped particle in the axial direction, as is known to be the case in the radial direction \cite{bera_2016,sil_2017,sil2020study}. These need to be carefully studied in future research, along with more detailed and intensive modeling of the dynamics of trapped particles that are confined using photophoretic forces generated by a multi-mode and a single mode beam profile. 

While the detailed theory of photophoretic trapping itself is not available yet, with only quasi-emperical models proposed, our experiments definitively confirm the significant advantage provided by single multi-mode fiber-based photophoretic traps in terms of robustness, portability, ease of use, inexpensiveness, and the facilitation of diverse applications towards simultaneous trapping and spectroscopy of aerosols/bio-aerosols, and other diverse applications. We hope to see exciting research in these directions in the near future.

% Experimental section

% Acknowledgements
\medskip
\textbf{Acknowledgements} \par %delete if not applicable))
The authors acknowledge IISER Kolkata, an autonomous institution funded by the Ministry of Education (MoE), Govt of India for funding and laboratory space. SS thanks CSIR, MoE for fellowship support.
% References
\medskip
\section{Appendix}
\subsection{Speckle Size Measurement}
We know that speckle is a random distribution of light field - consisting of a multitude of dark and bright spots resulting from destructive and constructive interference\cite{goodman1976some}. There are different speckle parameters such as mean speckle size, contrast, intensity and polarization etc\cite{piederriere2004scattering}. But here, we only consider the mean speckle size, defined as the average size of bright or dark spots present in the pattern\cite{goodman1976some}. Thus, in order to find out the mean speckle size, we need to measure the Wiener spectrum of the pattern, which is the average strengths of all possible spatial frequency components of the pattern\cite{goodman1976some}. This can be done by calculating the normalized autocovariance function of the intensity speckle pattern obtained in the observation or image plane (x,y). Further, this function can be considered as the normalized autocorrelation function of the intensity, which has a zero base, and its width provides a good measurement of the average width of a speckle\cite{piederriere2004scattering,hamarova2014methods}.

Let us consider, $I(i_1,j_1)$ and $I(i_2,j_2)$ to be the gray values of two pixel points in the image plane (i,j). Then, the intensity autocorrelation function is defined as, 
\begin{equation}
	R_I(\Delta i, \Delta j) = \langle I(i_1,j_1) I(i_2,j_2)\rangle
	\label{eq1}
\end{equation}
where, $\Delta i = i_1 - i_2$ and $\Delta j = j_1 - j_2$; and $\langle .. \rangle$ represents a spatial average. Now, for simplicity, we assume $i_2 = 0$, $j_2 = 0$ and consider $i_1 = i$ and $j_1 = j$, so the Eq.~\ref{eq1} can be written as, 
\begin{equation}
	R_I(\Delta i, \Delta j) = R_I(i,j)
	\label{eq2}
\end{equation}
So, the normalized autocorrelation function of the intensity ($C_I(i,j)$) can be expressed as, 
\begin{equation}
	C_I(i,j) = \frac{R_I(i,j) - {\langle I(i,j) \rangle}^2}{\langle I(i,j)^2 \rangle - {\langle I(i,j) \rangle}^2}
	\label{eq3}
\end{equation}
Again, according to the Wiener-Khinchin theorem, the power spectral density of the wide-sense-stationary random process is the Fourier transform of the corresponding autocorrelation function, which is again the square modulus of the Fourier transform of the signal (I(i,j)). Hence, we can write, 
\begin{equation}
	PSD_I(\nu_i,\nu_j) = FT[R_I(i,j)] = {| FT[I(i,j)] |}^2 
	\label{eq4}
\end{equation}
Here FT represents the Fourier Transform. So, using the Eq.~\ref{eq4}, the normalized auto-correlation function of the intensity ($C_I(i,j)$) [Eq.~\ref{eq3}] can be expressed as: 
\begin{equation}
	C_I(i,j) = \frac{{FT}^{-1}[{|FT[I(i,j)]|}^2] - {\langle I(i,j) \rangle}^2}{\langle I(i,j)^2 \rangle - {\langle I(i,j) \rangle}^2}
	\label{eq5}
\end{equation}
Thus, we determine the normalized auto-correlation intensity distribution ($C_I(i,j)$) of any given pattern image using the algorithm described by Eq.~\ref{eq5}.  

\subsection{Numerical estimation of Photophoretic Forces}
We now numerically estimate the value of the photophoretic forces and radiation pressure force experienced by the particle. Hence, we employ the analytical formula for photophoretic $\Delta T$  and $\Delta \alpha$ forces acting on a spherical particle provided by Rohatschek in his semi-empirical model of photophoretic forces\cite{rohatschek_1995}. Note that the quantitative analysis of photophoretic forces acting on a particle is quite complex, as many factors are involved with this force - such as pressure, different parameters of light (beam profile, intensity, wavelength of the laser, etc.), and most importantly, particle properties (i.e., particle size, morphology, thermal conductivity, absorptivity, etc.)\cite{horvath_2014}. Thus, only semi empirical estimates of the forces are available from the literature. However, the calculation of the photophoretic forces significantly depend on the Knudsen number ($k_n$) - defined as the ratio of the mean free path of the gas molecule ($\lambda$) and the size of the particle (a), $K_n = \frac{\lambda}{a}$ \cite{horvath_2014}. When $K_n > 1$ i.e., the particle size is considerably smaller than the mean free path of the gas molecules. This also includes conditions corresponding to very low pressure $(p \rightarrow 0)$ (mean free path of the gas molecules increases), where the free-molecular regime is applied for the calculation of photophoretic forces. For $K_n < 1$ i.e., when the particles size is much larger than the $\lambda$, or for  atmospheric and high pressures, the continuum regime can be applied. Since we trap particles in air at atmospheric pressure, where the mean free path of the air molecules are in the order of nanometers and the particle size is in the order of mirometers, $K_n << 1$ - hence the calculation of the photophoretic forces are based on the continuum regime\cite{horvath_2014}.

Photophoretic forces may be generated both by the difference in the surface temperature ($T_s$) and by variations in the thermal accommodation coefficient ($\alpha$) of the particle. In our case, we assume a spherical particle of radius $a$, heated to a certain temperature ($T_s$) due to laser radiation. When gas molecules having temperature $T_i$ ($<T_s$) are incident on the surface of the particle, they are reflected off the particle and reach a higher temperature $T_r$. This elevated temperature can be written using Knudsen's concept of energy transfer by individual gas molecules interacting with a hotter surface as,
\begin{equation}
	T_r = T_i + \alpha(T_s - T_i) 
	\label{eq6} 
\end{equation}
where, $\alpha$ is the thermal accommodation coefficient. Let us consider the temperature of the gas layer adjacent to the surface of the particle to be ($T_a$), which can be expressed in terms of $T_i$ and $T_r$ as,
\begin{equation}
	T_a = \frac{n_i T_i + n_r T_r}{(n_i + n_r)}
	\label{eq7} 
\end{equation}
where, $n_i$ and $n_r$ depicts the number density of the incident and reflected air molecules respectively, and the continuity at the surface signifies - $n_r \overline{c_r} = n_i \overline{c_i}$, so that, $n_r \sqrt{T_r} = n_i \sqrt{T_i}$, where $\overline{c} \ [= \sqrt{8RT/\pi M}]$ is the mean velocity of the gas molecules. Substituting the above equation in Eq. \ref{eq7},  and further approximating the geometric mean by the arithmetic mean, the expression of $T_a$ becomes,
\begin{equation}
	T_a = \sqrt{T_i \ T_r} = \frac{T_i + T_r}{2}
	\label{eq8}
\end{equation}
As we consider a spherical particle, we assume that the distribution of the difference in temperature ($\Delta T_s$), and accommodation coefficient ($\Delta \alpha$) are rotationally symmetric, where $T_s$ is measured about the direction of incident light, and $\alpha$ about an axis fixed to the particle. Besides, the temperature of the gas layer adjacent to the surface of the particle ($T_a$) is also assumed to have rotational symmetry. Thus, all the relevant quantities $T_s$, $\alpha$ and $T_a$ can be expanded in terms of the Legendre polynomial $P_n (cos\theta)$, so that the surface temperature ($T_s$) of the particle can be written as, 
\begin{equation}
	T_s = T_\infty + \sum_{n = 0}^{\infty} A_n P_n (cos\theta) = {T_s'} + A_1 cos\theta + . . .
	\label{eq9}
\end{equation} 
where, $T_\infty$ denotes the temperature of the gas far from the sphere and ${T_s'} = T_\infty + A_0$. \\
Similarly, $\alpha$ can be expressed as,
\begin{equation}
	\alpha = \sum_{n = 0}^{\infty} a_n P_n (cos\theta) = a_0 + a_1 cos\theta + . . .
	\label{eq10}
\end{equation} 
And, the gas temperature next to the surface $T_a$ can be represented as,  
\begin{equation}
	T_a = T_\infty + \sum_{n = 0}^{\infty} B_n P_n (cos\theta) = {T_a'} + B_1 cos\theta \\
	\label{eq11}
\end{equation}
where,  ${T_a'} = T_\infty + B_0$.

In the following section, we calculate the photophoretic $\Delta T_s$ and $\Delta \alpha$ forces where for the $\Delta T_s$ force, $\alpha$ is assumed to be constant, while for the $\Delta \alpha$ force, we consider a constant average surface temperature of the particle ($\overline{T_S}$). 

\subsubsection{Calculation of Photophoretic $\Delta T_s$ force}
We consider a spherical particle of radius $a$, which is highly light-absorbing, and has very low thermal conductivity $k_p$. The particle is placed in an air medium at normal atmospheric pressure ($p$), and has a molecular weight $M$ and viscosity $\eta$. The particle is illuminated by intense laser light in a direction opposite to gravity. Due to the high absorptivity of the particle at the operating wavelength of laser light, the surface facing the illumination source is warmer than the opposite side - resulting in a force in the direction of propagating light, termed as  positive photophoresis force. The direction of this force is solely determined by the incident laser light direction - and is independent of the particle orientation, and hence called space-fixed. As the temperature difference across the particle surface causes the force, which is directed longitudinally, it is also named as longitudinal photophoresis force $(\Delta T_s)$ force.

However, for the continuum regime, the photophoretic $\Delta T_s$ force is derived from the 'thermal creep' flow around the particle and the resulting viscous forces. In this regime, it is generally considered in the literature that the gas molecules adjacent to the particle have a similar temperature as the particle's surface. As shown in Fig.~\ref{Fig1}, it is expected that for such a highly absorbing particle, the left side has a higher temperature $T_{s1}$ compared to the right side $T_{s2}$, in accordance with the illumination on the particle. Thus, the air molecules impacting the left (hot) side are faster than those impacting the right (cold) side.

\begin{figure}[h]
	\centering
	%\captionsetup{justification=centerlast}
	\includegraphics[scale=0.60]{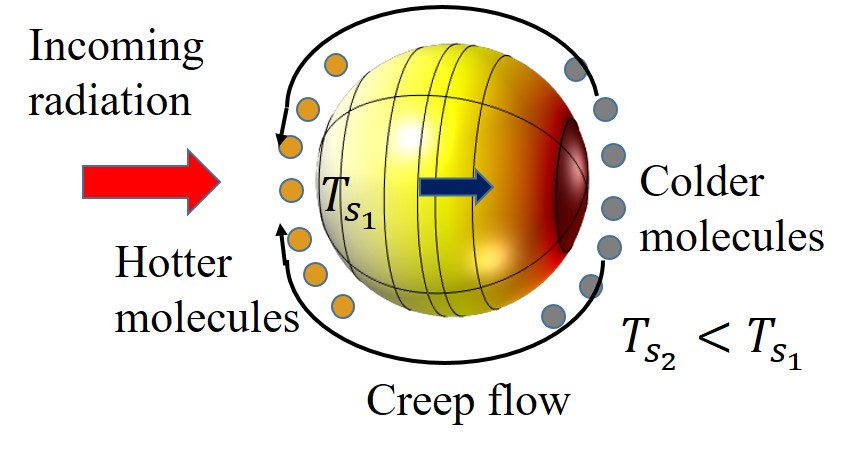}
	\caption{Photophoretic $\Delta T_s$ force acting on a particle arising from thermal creep flow}
	\label{Fig1}
\end{figure}

Now, due to the thermal accommodation, a symmetric velocity distribution of the gas molecules is obtained with respect to the perpendicular direction of the surface. Hence, the warmer molecules coming from the left side transfer larger momentum to the particle along in the right direction compared to the colder molecules to the left - resulting in the particle experiencing a net force that is directed from the hot to the cold side. The air molecules correspondingly lose the same amount of momentum, so that a creeping flow around the particle from the cold side to the hot side occurs simultaneously. Now, the thermal creep velocity can be written as\cite{rohatschek_1995},

\begin{equation}
	u_s = \kappa \frac{\eta}{\rho_{air} T} \frac{dT}{ds}
	\label{eq12}
\end{equation}

where, $u_s$, $\eta$, and $\rho_{air}$ are denoted the tangential velocity of the gas, the dynamic viscosity, and the mass density, respectively. $T$ and $\frac{dT}{ds}$ are the temperature and tangential temperature gradient in the gas adjoining the particle surface, respectively. T is written as $T_a$, $\kappa$ is defined as the thermal creep coefficient, which is connected to the momentum accommodation coefficient having value $\kappa = 1.14$. The net photophoretic force can then be obtained by integrating the stress components in the z-direction over the surface\cite{rohatschek_1995} (Eq.\ref{eq12}), as,
\begin{equation}
	F = 4\pi \kappa \frac{\eta^2}{\rho T_a} A_1  =  4\pi \kappa \frac{R \eta^2}{M p} A_1,
	\label{eq13}
\end{equation}
where, $A_1$ is the first order Legendre coefficient corresponding to the particle's surface temperature $T_s$. We can rewrite Eq.\ref{eq13} in terms of  two other parameters $D$ and $p*$, as:
\begin{equation}
	F = 4\pi \kappa \frac{\eta^2}{p} \left(\frac{R}{M}\right) A_1 = 4\pi \kappa \frac{\eta^2}{p} \left(\frac{\overline{c}^2 \pi}{8 T}\right) A_1 = 2 D \ \frac{p*}{p} \ a A_1  
	\label{eq14}
\end{equation}
where $D$ denotes a constant, determined entirely by the state of the gas and $p*$ is the characteristic pressure that depends on particle radius, and is expressed as,
\begin{eqnarray}
	D &=& \frac{\pi}{2} \sqrt{\frac{\pi}{3} \kappa} \  \frac{\overline{c} \eta}{T_a} \\
	p* &=& \frac{1}{2} \sqrt{3 \pi \kappa} \  \overline{c} \eta \ \frac{1}{a}
\end{eqnarray}
However, the parameter $A_1$ can be calculated either by knowing the value of the surface temperature difference of the particle or the laser irradiance. Here we consider only the known surface temperature difference of the particle ($\Delta T_s$) that can be written as\cite{rohatschek_1995},
\begin{equation}
	T_s = \overline{T_s} + \frac{1}{2} \Delta T_s cos\theta 
	\label{eq17}
\end{equation}
Again, comparing  Eq.~\ref{eq17} with Eq.~\ref{eq9}, we obtain, 
\begin{equation}
	A_1 = \frac{1}{2} \Delta T_s
	\label{eq18}
\end{equation} 
Hence, the final form of the photophoretic $\Delta T_s$ force can be found by putting the expression of Eq. \ref{eq18} for $A_1$ into the Eq. \ref{eq9}, as: 
\begin{equation}
	F = D  \ \frac{p*}{p} \ a \Delta T_s
	\label{eq19}  
\end{equation} 

\subsubsection{Estimation of the Photophoretic $\Delta \alpha$ force}
The photophoretic $\Delta \alpha$ force arises due to
the difference in accommodation coefficient of the particle surface, which might be caused due to differences in surface roughness or composition of the particle. For the calculation of $F_{\Delta \alpha}$ we consider that the particle's surface has two different thermal accommodation coefficient values $\alpha_1$ and $\alpha_2$, where $\alpha_1 > \alpha_2$, and assume that the particle has an average surface temperature ($T_s$) which is hotter than the surrounding air molecules. Note that a higher value of the thermal accommodation coefficient signifies a higher heat transfer rate. Thus, the surface of the particle with a higher value of $\alpha$ transfers more heat to the air molecules compared to the other surface having a lower accommodation coefficient value - resulting in the particle experiencing a thrust in the direction of the high accommodation ($\alpha_1$) surface to the low accommodation surface ($\alpha_2$). However, the photophoretic $\Delta \alpha$ force calculation is based on a common photophoretic function ($\phi$), and an estimation of $B_1$ - which is a first-order Legendre coefficient of temperature distribution in the gas layer next to the surface - arises due to the difference in the accommodation coefficient ($\alpha \neq 0 $), and not by a difference in the surface temperature($A_1 = 0$). So, the photophoretic $\Delta \alpha$ force can be written as,
\begin{equation}
	F_{\Delta \alpha} = \phi * B_1 
	\label{eq20}
\end{equation}
This photophoretic function $\phi$ is entirely dependent on the gas temperature, and hence it covers both $\Delta T_s$ and $\Delta \alpha$ forces and can be expressed as\cite{rohatschek_1995, reed1977low}
\begin{equation}
	\phi = D \frac{2}{\left(\frac{p}{p^*} + \frac{p^*}{p}\right)} a 
	\label{eq21}
\end{equation}
The next step is to calculate the $B_1$ value, and for that, the molecular energy transfer in the Knudsen layer needs to be considered. As the gas temperature next to surface $T_a$ depends on the accommodation distribution values of the particle surface, a relationship between $B_1$ and $a_1$ - where the latter is the first order Legendre coefficient of the accommodation coefficient - can be found out using Eqs. (\ref{eq6}), (\ref{eq8}),(\ref{eq10}) and (\ref{eq11}), assuming that $T_s = \overline{T_s}$.
\begin{eqnarray*}
	\hspace{70pt} 2T_a &=& \alpha (\overline{T_s} - T_i) + 2T_i \hspace{10pt} [Putting \ Eq. \ (\ref{eq8}) \ into \ Eq. \ (\ref{eq6}) \ ] \\
	\Rightarrow T_a &=& \frac{\alpha}{2} (\overline{T_s} - T_i) + T_i \\
	\Rightarrow T_a &=& \frac{1}{2} (a_0 + a_1 cos \theta) (\overline{T_s} - T_i) + T_i  \hspace{10pt} [From \ Eq. \ (\ref{eq11}) \ \alpha = a_0 + a_1 cos\theta \ ] \\
	\Rightarrow T_a &=& \left[\frac{1}{2}(\overline{T_s} - T_i)a_0 + T_i \right] + \frac{a_1}{2} (\overline{T_s} - T_i) cos \theta 
\end{eqnarray*} 
Comparing this with Eq.~(\ref{eq11}) $[T_a = \overline{T_a} + B_1 cos \theta]$, we obtain,
\begin{equation}
	B_1 = \frac{a_1}{2} (\overline{T_s} - T_i) 
	\label{eq22}
\end{equation}
Again, the value of coefficient $a_0$ and $a_1$ can be obtained as,
\begin{equation}
	a_0 = \overline{\alpha}; \hspace{10pt} a_1 = \frac{3}{4}\Delta \alpha
	\label{eq23}  
\end{equation}
where, $\overline{\alpha} = \frac{\alpha_1 + \alpha_2}{2}$, and  $\Delta \alpha = \alpha_1 - \alpha_2$. Plugging Eq.(\ref{eq23}) into (\ref{eq22}), the value of $B_1$ becomes,
\begin{equation}
	B_1 = \frac{3}{8} (\overline{T_s} - T_i) \Delta \alpha
	\label{eq24} 
\end{equation}
So, the photophoretic $\Delta \alpha$  force can be obtained by putting the expression of $\phi$ [Eq.\ref{eq21}] and $B_1$ [Eq.\ref{eq24}] into the Eq.\ref{eq20} and the result is
\begin{equation}
	F_{\Delta \phi} = \phi * B_1 =\frac{3}{4} D \frac{1}{\left(\frac{p}{p^*} + \frac{p^*}{p}\right)} a  (\overline{T_s} - T_i) \Delta \alpha 
	\label{eq25} 
\end{equation} 
Thus, we use the expression of Eq.~\ref{eq19} and Eq.~\ref{eq25} for the determination of the photophoretic $\Delta T$ and $\Delta \alpha$ forces, respectively. 
% Use the following code if you wish to generate your bibliography with BibTeX;
% replace the string "MSP-template" below with the name(s) of
% the BibTeX data base(s) you want to use.
% The resulting bibliography-output (the content of the .bbl file)
% must be pasted back into this file before submission.
% Please also include your BibTeX data base file(s) in your submission
% so that we can re-run BibTeX if necessary.
%
\bibliographystyle{apsrev4-2}
%\bibliography{Reference_MM}

\end{document}